# Test Planning for Mixed-Signal SOCs with Wrapped Analog Cores

Anuja Sehgal, Fang Liu, Sule Ozev and Krishnendu Chakrabarty

Department of Electrical & Computer Engineering

Duke University, Durham, NC 27708, USA

{as,sule,krish}@ee.duke.edu

## Abstract

*Many SOCs today contain both digital and analog embedded cores. Even though the test cost for such mixed-signal SOCs is significantly higher than that for digital SOCs, most prior research in this area has focused exclusively on digital cores. We propose a low-cost test development methodology for mixed-signal SOCs that allows the analog and digital cores to be tested in a unified manner, thereby minimizing the overall test cost. The analog cores in the SOC are wrapped such that they can be accessed using a digital test access mechanism (TAM). We evaluate the impact of the use of analog test wrappers on area overhead and test time. To reduce area overhead, we present an analog test wrapper optimization technique, which is then combined with TAM optimization in a cost-oriented heuristic approach for test scheduling. We also demonstrate the feasibility of using analog wrappers by presenting transistor-level simulations for an analog wrapper and a representative core. We present experimental results on test scheduling for an ITC'02 benchmark SOC that has been augmented with five analog cores.*

## 1 Introduction

Advances in semiconductor technology are contributing to the increasing complexity of system-on-chip (SOC) integrated circuits. Many SOCs in use today are mixed-signal circuits containing both digital and analog embedded cores [1, 2]. There are enormous costs associated with the testing of mixed-signal SOCs. The cumulative test cost of an SOC has three main components: (i) the cost of the Automatic Test Equipment (ATE); (ii) the cost of silicon area overhead due to the on-chip test hardware; (iii) the cost due to test application time. In order to reduce the overall test cost of mixed-signal SOCs, all of the above components of the test cost should be minimized.

Most prior research on test cost reduction for SOCs has focused on digital SOCs. However, since the test cost of a mixed-signal SOC is much higher than that of digital SOCs [3] and many SOCs today have significant analog content, there is a need for efficient test methodologies that can handle mixed-signal SOCs and reduce their test cost. Many consumer electronics products, such as MP3 players, PDAs, and audio receivers contain a small number of analog cores that operate in the low to mid-frequency range; these cores are embedded in an SOC together with a large number of digital cores. Consumer electronics products belong to a high volume, low profit-margin domain, where reducing test cost is of prime importance. Modular testing of embedded cores in SOCs is being increasingly advocated to simplify test generation, enhance test reuse, and reduce test cost [4]. Test wrappers are used to isolate a core, while test access mechanisms (TAMs) transport test patterns and test responses between SOC pins and core I/Os.

In [5], preliminary work was done on the use of analog test wrappers to eliminate the need for expensive mixed-signal testers and allow a unified test approach for both digital and analog cores. The analog test wrappers can be used for low-frequency applications that require analog tests in the audio frequency range. These wrappers allow an unified testing of the digital and analog cores in an SOC, thereby reducing test application time. In this paper, we improve upon [5] in the following ways:

- We propose a new resource optimization technique that reduces the overall area and routing overhead by using shared test wrappers for the time-multiplexed testing of analog cores.

- We propose a test planning method that combines a previously developed TAM optimization approach [6] with the new resource optimization approach. It leads to a TAM architecture that is efficient in terms of area, routing costs, and overall test time.

- We implement analog wrappers in a $0.5 \mu m$ AMI technology and present transistor-level simulation results.

Analog wrappers can contribute significantly to area overhead. Hence, if the number of test wrappers is reduced, the area overhead is also reduced. In the proposed resource optimization technique, we share test wrappers for time-multiplexed testing of analog cores. This approach reduces the area overhead due to the analog wrappers; however, it can potentially increase the testing time for the SOC.

The proposed cost optimization approach evaluates judiciously chosen combinations of shared analog wrappers, and chooses the best wrapper architecture for the analog cores in terms of overall test cost of the SOC. A pruning technique based on lower bounds on the test time, area overhead, and routing overhead, is used to reduce the number of wrapper combinations that are evaluated. Hence this approach is computationally inexpensive. In the absence of mixed-signal SOC benchmarks, we present results for a "mixed-signal SOC" that has been crafted by adding five analog cores to a digital SOC from the ITC'02 SOC test benchmarks [7]. These results demonstrate that a significant reduction in the overall test cost can be achieved using the proposed approach.

The rest of this paper is organized as follows. Section 2 presents a review of relevant prior work. In Section 3, we detail the analog wrapper optimization approach, which is followed by a description of the cost-optimization approach in Section 4. In Section 5, we describe the implementation of the wrapper architecture and present results for one of the analog core tests. In Section 6, we present experimental results for an ITC'02 benchmark SOC. Finally, we conclude the paper in Section 7.

## 2 Review of Prior Work

In the analog testing domain, research has primarily focused on defining core-level measurement and test methods. Attempts have also been made to reduce the overall test time for analog circuits. In [8, 9], only a subset of parameters are tested, which are selected based on parameter correlations. Automated generation of test stimuli has also been used as a means to reduce test time of analog circuits [10].

---

[1] This research was supported in part by the Semiconductor Research Corporation under contract no. 2004-TJ-1176.

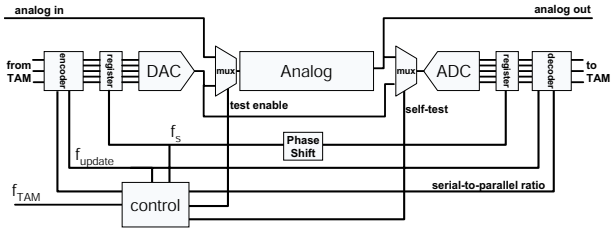

**Figure 1.** Block diagram of the analog test wrapper [5].

Significant research has been done to eliminate the need for an expensive analog or mixed-signal ATE. The use of on-chip data converters, proposed in [11, 12], obviates the need for expensive analog testers. Several BIST techniques have also been proposed for mixed-signal blocks that cannot be directly tested by an ADC-DAC pair. Such BIST techniques target either data converters themselves [16, 17, 18, 15] or PLLs [21, 20, 19]. Recently, in [5], test wrapper design for analog cores has been developed to obviate the need for mixed-signal testers. The analog test wrappers contain on-chip data converters that convert analog cores into virtual digital cores. Thus, the wrapped analog cores can be tested in a unified manner with the digital cores on a digital TAM. The test wrappers are reconfigurable for different data resolutions and frequencies. The reconfigurability allows the use of the wrapper for a variety of analog tests that may potentially differ in TAM width and sampling frequency requirements. The on-chip implementation of the data converters can be used for a wide range of low frequency application.

In [5], however, the impact of test wrappers on overall area overhead of the SOC was not considered. Also, an on-chip implementation of the wrappers was not done to evaluate their feasibility.

Figure 1 shows a block diagram of the analog wrapper proposed in [5]. The control and clock signals generated by the test control circuit are highlighted. The registers at each end of the data converters are written and read in a semi-serial fashion depending on the frequency requirement of each test. The digital test control circuit selects the configuration for each test. This configuration includes the divide ratio of the digital TAM clock, the serial to parallel conversion rate of the input and output registers of the data converters, and the test modes. The test modes of the wrapper include a normal mode, a self-test mode and a core-test mode.

With the conversion of the analog cores into virtual digital cores, TAM optimization techniques can be used to optimize a digital TAM architecture for the testing of digital and analog cores of mixed-signal SOCs.

## 3 Analog Wrapper Optimization

In Section 2, we discussed the analog test wrapper design proposed in [5]. The ADC-DAC pair, together with the encoder-decoder pair, forms the predominant part of the analog test wrapper. The encoder and decoder allow the wrapper to be reconfigured for a set of different tests. In our proposed approach, we exploit this feature of reconfigurability to optimize the resource utilization and reduce the area overhead cost.

We propose that an analog test wrapper be designed such that it can support testing of more than one analog core multiplexed in time from one test to another. In our proposed approach we use the reconfigurability feature of the analog wrappers to allow the test of multiple analog cores, using a single wrapper, thereby reducing the overall area overhead significantly. The design proposed in [5] can be easily modified to accommodate this feature. Figure 2 illustrates two analog cores sharing test wrappers (only the ADC-DAC pair of the wrappers are shown for the purpose of illustration). The time-multiplexed testing of the cores can be ensured by the use of multiplexers. Although the use of analog multiplexers may result in additional parasitic noise, the use of analog multiplexers is an accepted practice in analog testing, and design methods exist to alleviate the noise problem [22, 23, 24, 25]. The sizes of the encoder, decoder, and the ADC-DAC pair in a shared analog wrapper are determined such that they can satisfy the requirements of all the cores sharing the wrapper. The resolution of the ADC-DAC pair in the proposed shared analog wrapper is selected to be the maximum of the ADC-DAC resolution requirements of all the analog cores sharing the wrapper. Similarly, the encoder and decoder are designed for the test with the largest TAM width requirement. The encoders and decoders can be configured to test any of the analog cores. However, a module that requires high-speed and low-resolution data converters cannot share its wrapper with a module that requires high-resolution and low-speed data converters. It may not be feasible to satisfy the requirements of high-speed and high-resolution with reasonable overhead.

Wrapper sharing results in a certain routing overhead that needs to be accounted for. For analog cores that are separated by a large distance, sharing is less advantageous since the routing overhead will be high. In this work, we evaluate the area overhead due to analog test wrappers as follows. The routing overhead is considered to be a percentage of the wrapper architecture's area overhead. This percentage depends on the relative on-chip location of the analog cores. Typically this location is determined by the functional proximity between the analog core and other cores in the system. Thus, an approximate idea about the proximity of analog wrappers can be obtained prior to layout. The area overhead is estimated as the ratio of the area overhead due to sharing, to the area overhead if there is no sharing of wrappers. When there is no sharing of wrappers between cores, the area overhead is maximum. The area overhead due to test wrappers can be expressed as:

$$\mathcal{C}_A = \frac{\sum_i^{N_w}(1+d_i) \cdot Amax_i}{\sum_j^N A_j} \times 100 \qquad (1)$$

where,
$N_w$: number of analog wrappers used;
$N$: the number of analog cores;
$d_i$: the routing overhead for shared wrapper $i$;
$A_j$: area overhead of analog wrapper $j$;
$Amax_i$: maximum of the individual wrapper area overheads of the cores for the shared wrapper $i$.

The cost function defined above is used for preliminary cost analysis. Using the above estimate, it is possible to determine the relative cost of the different sharing combinations among the various analog cores. The routing overhead of a wrapper that serves $n$ cores is defined as $d_i = (n-1) \times p$, where $0 < p \leq 1$ is a factor proportional to the cumulative distance of the $n$ cores from each other. In this work, without loss of generality, we have considered a representative value of $p = 0.15$ to illustrate the approach. Thus, wrappers that serve only one core have a routing overhead of $d_i = 0$. Note that $C_A$ should always be lower than 100. The sharing combinations that exceed the overhead of the no-sharing case should not be considered.

In order to avoid potential resource conflicts, it is imperative that the tests for cores that share a wrapper do not not overlap in time in

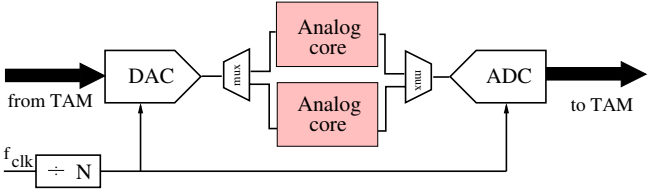

**Figure 2.** Shared test wrapper for analog cores.

| $n$ | $C$ | $\mathcal{C}_A$ | $t_{LB}$ | $n$ | $C$ | $\mathcal{C}_A$ | $t_{LB}$ |
|---|---|---|---|---|---|---|---|
| 4 | {A,C} | 83.0 | 68.5 | 3 | {A,D,E} | | 31.4 |
| | {C,D} | | 56.0 | 2 | {A,B,C,D} | | 98.7 |
| | {C,E} | | 48.3 | 2 | {A,B,C,E} | 49.0 | 91.1 |
| | {A,B} | | 42.7 | 2 | {A,C,D,E} | | 78.6 |
| | {A,D} | | 30.2 | 2 | {A,B,D,E} | | 52.8 |
| | {A,E} | | 22.6 | 2 | {A,B,C}{D,E} | 46.0 | 89.8 |
| | {D,E} | | 10.1 | 2 | {A,C,D}{B,E} | | 77.3 |
| 3 | {A,B,C} | 66.0 | 89.8 | 2 | {A,C,E}{B,D} | | 69.7 |
| | {A,C,D} | | 77.3 | 2 | {A,D,E}{B,C} | | 68.5 |
| | {A,C,E} | | 69.7 | 2 | {C,D,E}{A,B} | | 57.2 |
| | {A,B,D} | | 57.2 | 2 | {A,B,E}{C,D} | | 56.0 |
| | {C,D,E} | | 51.6 | 2 | {A,B,D}{C,E} | | 51.6 |
| | {A,B,E} | | 43.9 | 1 | {A,B,C,D,E} | 100 | 100 |

$n$: Number of wrappers; $C$: Combination of cores that share a wrapper.

**Table 1.** Area overhead costs for all combinations of wrapper sharing.

the test schedule. Thus, we constrain the TAM optimization procedure such that the tests for cores sharing the same wrapper are scheduled serially in time. In this way, the total test time usage of the test wrapper is the sum of the test times of the analog cores that share the wrapper. A lower bound on the overall test time of all the analog cores can now be calculated as the maximum of the usage of every analog test wrapper, i.e., if three analog test wrappers are used to test all the analog cores, then a lower bound $t_{LB}$ on the test time is the maximum of the test time usage of the three analog test wrappers. Table 1 shows the $\mathcal{C}_A$ values for all the combinations of sharing between the five analog cores considered in the experimental setup. The normalized lower bound for each case is also presented; these values have been normalized to the maximum lower bound. A detailed description of the five analog cores labeled $A$ to $E$ is presented in Section 6. (Since Core $A$ and Core $B$ have identical tests, only unique combinations for Core $A$ are presented.)

## 4 Test Cost Optimization

In this section, we define the test cost minimization problem for a given TAM width $W$. The objective is to minimize the test cost in terms of test application time and the area overhead.

We use a previously developed TAM optimization technique, based on rectangle packing [6], to obtain the test application time for an SOC. Unlike the approach described in [5], this approach exploits the disparity in the TAM width requirements of digital and analog cores to reduce the overall test time of the SOC. The TAM width requirements of an analog core are usually much smaller than that of most digital cores [5]. Moreover, their testing time does not reduce with an increase in the number of digital TAM wires allocated for them. For digital cores, there exists a "staircase variation" of testing time with TAM width [13], hence their testing time can be reduced with an increase in the TAM width. Thus, there is often a substantial disparity between the TAM width requirements of digital and analog cores. As a result, when analog cores are tested serially with digital cores on the same TAM partition, the analog cores do not use all the TAM wires. Consequently the overall time taken to test the SOC is not optimized.

We therefore use a TAM optimization approach, based on a flexible-width TAM architecture, that can handle digital and analog cores in a unified manner, yet bridge the gap in TAM width requirements of digital and analog cores.

The test cost for a given SOC-level TAM width $W$ can be minimized as follows. The total test cost is expressed as

$$\mathcal{C}(W) = \omega_T \cdot \mathcal{C}_T(W) + \omega_A \cdot \mathcal{C}_A \qquad (2)$$

where $\omega_T$ is the cost weighting factor for the test application time $\mathcal{C}_T$, and $\omega_A$ is the cost weighting factor for the area overhead cost $\mathcal{C}_A$. The weighting factors are defined such that $\omega_T + \omega_A = 1$. The cost of test application time is expressed as $\mathcal{C}_T(W) = 100 \times T(W)/T_m(W)$, where $T_m(W)$ is the test time of the SOC when all the analog cores share a single analog wrapper. This case represents the most constrained scenario for test scheduling, hence for any given TAM width, it is likely to yield the highest test time. Essentially, $\mathcal{C}_T$ is the test time normalized to the maximum possible test time. The TAM optimization procedure is used to obtain the value of $T_m(W)$ for a given TAM width $W$. The area overhead cost includes the cost of the analog core wrappers and the routing overhead of shared wrappers as explained in Section 3. Both the costs $C_T$ and $C_A$ have been defined to have values between 1 and 100.

Now, the problem of minimizing the overall test cost of an SOC can be stated as follows.

**Problem $P_{cost}$**: Given the test data parameters for the digital cores, the testing time in clock cycles and the core-level TAM widths for the analog cores, the total SOC-level TAM width $W$, and the test time cost and area overhead cost weights $\omega_T$ and $\omega_A$ respectively, determine (i) the wrapper design for digital cores, (ii) the groups of analog cores that share analog wrappers, (iii) the TAM width for each core and test schedule for the SOC, such that the total number of TAM wires utilized at any moment does not exceed the overall TAM width $W$, and the total cost $\mathcal{C}(W) = \omega_T.\mathcal{C}_T(W) + \omega_A.\mathcal{C}_A$ is minimized. □

The *Design_wrapper* algorithm from [13] is used to design the wrappers for digital cores. Next, the grouping of the analog cores is determined, such that the analog cores grouped together share the same analog test wrapper. Finally, the TAM optimization approach is used to to determine a test schedule for the digital and analog cores.

Depending on the specified weights $\omega_T$ and $\omega_A$, the analog cores can be grouped such that they share analog wrappers and the overall cost of the wrappers is minimized. The degree of sharing is dictated by the weighting factors in the cost function. If $\omega_T > \omega_A$, the test time is given more weightage in optimization. In this case, the degree of wrapper sharing may be chosen such that the area overhead cost reduction is compromised to achieve better test times. Similarly if $\omega_A > \omega_T$, the degree of sharing is chosen such that the area overhead minimization has priority over test time minimization.

One approach for solving problem $P_{cost}$ is to evaluate the overall cost $\mathcal{C}_T$ for every possible configuration of shared analog wrapper (as presented in Table 1) for a given TAM width $W$ and weights $\omega_T$ and $\omega_A$. This exhaustive approach requires the TAM optimization procedure to be run for every combination of analog cores to obtain the $\mathcal{C}_T(W)$ values. This is computationally expensive for a larger problem instance with many analog cores since the number of distinct combinations increases exponentially with the number of analog cores.

We propose a heuristic approach that scales well with the increase in the number of analog cores and provides a near optimal result. We use a pruning technique based on the area overhead costs $\mathcal{C}_A$ and analog test time lower bounds $t_{LB}$, which are available prior to cost op-

**Procedure** $Cost\_Optimizer(\omega_T, \omega_A, W)$

1. Group combinations having the same degree of sharing to form $G = G_1 \cup G_2 \cup \cdots \cup G_n$;
2. **for** $i := 1$ **to** $n$ **do**
3.   **for** $j := 1$ **to** $(|G_i|)$ **do**
4.     Evaluate $Ctemp(G_i(j)) := \omega_T.t_{LB}(G_i(j)) + \omega_A.C_A(G_i(j))$;
5.   **od**;
6. **od**;
7. **for** $i := 1$ **to** $n$ **do**
8.   Select $k \in \{j \mid Ctemp(G_i(j)) = max_{1 \leq x \leq n} Ctemp(G_i(x))\}$;
9.   Run $TAM\_Optimizer$ procedure to get the test times $T_i(W)$ of element $G_i(k)$; **od**;
10. $T_m(W) := max_{1 \leq x \leq n}(T_x(W))$
11. **for** $i := 1$ **to** $n$ **do**
12.   Evaluate cost $C_i(W)$ for selected element;
13. **od**;
14. Select $k \in \{j \mid C_{min} = min_{1 \leq x \leq n} C_x(W)\}$;
15 **for** $i := 1$ **to** $n$ **do**
16.   **if** $C_i - C_{min} > \delta$, eliminate group $i$;
17. **od**;
18. Evaluate all elements of the groups that have not been eliminated;
19. Return the element that results in the smallest $C(W)$;

**Figure 3.** Pseudocode for procedure $Cost\_Optimizer$.

timization. Figure 3 details the pseudocode for the proposed heuristic procedure $Cost\_Optimizer$.

First (in Line 1), all the combinations of analog cores sharing test wrappers are grouped by their degree of sharing. i.e., combinations that have the same area overhead cost ($C_A$) are grouped together. All the groups together form a set $G$. The goal is to be able to eliminate an entire group without having to do complete evaluation. Complete evaluation for a combination entails finding a test schedule by using the TAM optimization procedure.

The next step (line 4) is to estimate preliminary costs for every combination based on area overhead, cost weights, and the lower bounds on analog test times. We calculate the preliminary costs $Ctemp$ for every combination as:

$$Ctemp(G_i(j)) = \omega_T.t_{LB}(G_i(j)) + \omega_A.C_A(G_i(j)). \quad (3)$$

where, $G_i(j)$ is combination $j$ of group $G_i$.

Based on the $Ctemp$ values, the combination/element that has the smallest $Ctemp$ values is chosen from every group $i$ (Line 8). Next, the *TAM_Optimizer* procedure is used to evaluate the $C_T$ values of the chosen elements of each group. These values are used to determine the $C_i(W)$ value for the chosen elements. The group with the minimum cost $C_{min}$ is not eliminated. Next, any group that satisfies the elimination criteria (i.e., $C_{min} > C_i + \delta$) is eliminated. The elimination criteria can be relaxed by making the threshold $\delta$ larger.

## 5 Analog Wrapper Implementation

We next present implementation details of the analog test wrapper and demonstrate its functionality by applying a test to a wrapped analog core. We design the wrapper using an 8-bit DAC-ADC pair. All simulations and layout are done in a $0.5\mu m$ process technology.

The implementation of the ADC and DAC in a wrapper is critical to the performance and area-overhead of the wrapper. We use a modular pipelined architecture for the 8-bit ADC [14], using two 4-bit flash ADCs and one 4-bit DAC. Figure 4(a) shows a block diagram of the ADC. The modular architecture of the ADC reduces the area overhead significantly. An $N$-bit flash ADC requires $2^N$ comparators, thus an 8-bit flash architecture typically requires 256 comparators. In contrast, the modular approach needs only 32 comparators. The comparators are the primary contributors to the overall area of the ADC. Similarly, we use a modular voltage-steering 8-bit DAC architecture [14], which is constructed from two 4-bit DACs. Figure 4(b) shows a block diagram of the 8-bit DAC. This modular approach reduces the number of resistors used by a factor of 8. Although the modular approach also adversely impacts the speed of operation of the data converters, it does not prevent us from achieving our desired performance for the low-speed applications that are being targeted here.

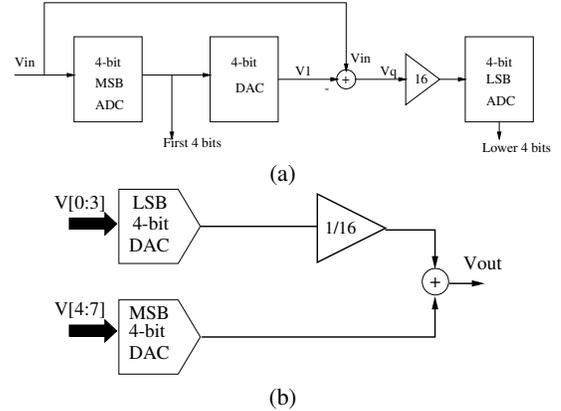

**Figure 4.** (a) Block diagram of a modular 8-bit ADC [14]; (b) Block diagram of a modular 8-bit DAC.

To demonstrate the accuracy of using digital test patterns to test wrapped analog cores, we apply a cut-off frequency test $f_c$ to analog core A (a detailed description of the core and its tests is presented in Table 2 of Section 6). The core is tested for cut-off frequency by applying a multi-tone signal. The frequency spectrum of the resulting signal is used to extrapolate the cut-off frequency of the filter. We compare the frequency spectrum obtained without using a wrapper and doing a direct analog test to that of the test responses obtained from the wrapped analog core. Figure 5 shows the HSPICE simulation results for the two scenarios. The error in the response from the wrapped analog core is approximately 5%. This error can be reduced further by using a more frequencies in the input signal; for the purpose of illustration, we have chosen an input with only three frequencies. The frequency spectrum is obtained by post-processing the transient analysis data obtained from the simulations. The system clock frequency is 50MHz and the sampling frequency of the input signal is 1.7MHz. The number of samples used is 4551. The supply voltage used is 4V.

We have also implemented a test chip for testing and characterizing an 8-bit analog wrapper. Its area in the $0.5\mu m$ process is only $0.02mm^2$. Preliminary comparison with an industrial core implemented in $0.12\mu m$ technology indicates that the wrapper, even though it is implemented in $0.5\mu m$ technology, is only one-eight the size of the core. We expect this ratio to be significantly smaller ($< 1/40$) if the wrapper is implemented in the same technology as the core. In this work, we have not considered the overhead of testing the ADC and DAC in the wrapper. Efficient BIST techniques can be used for testing the data converters [16, 17, 18] in the self-test mode of the wrapper.

## 6 Experimental Results

In this section, we present the description and specifications of the analog cores used in our mixed-signal SOC. Next, we study the impact of shared test wrappers on the overall SOC test cost. We also present a comparison of the cost-optimization heuristic with the exhaustive cost-optimization approach.

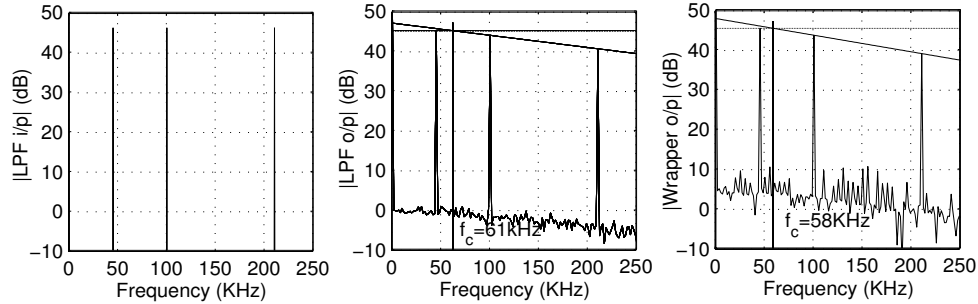

**Figure 5.** (a) Frequency spectrum of the applied analog test; (b) Frequency spectrum of the analog response of the core; Frequency spectrum of the response of the wrapped analog core.

| Test | $f_{min}$ | $f_{max}$ | $f_s$ | $T$ | $w$ |
|---|---|---|---|---|---|
| Cores $A$ & $B$: I-Q transmit ||||||
| $G_{pb}$ | 50kHz | 50kHz | 1.5MHz | 50,000 | 1 |
| $f_c$ | 45kHz | 55kHz | 1.5MHz | 13,653 | 4 |
| $G_{1MHz}$ & $G_{2MHz}$ | 1MHz | 2MHz | 8MHz | 12,643 | 2 |
| $IIP_3$ | 50kHz | 250kHz | 8MHz | 26,973 | 2 |
| $V_{off_{DC}}$ | DC | DC | 10kHz | 700 | 1 |
| $\Phi_{off}$ | 200kHz | 400kHz | 15MHz | 32,000 | 4 |
| Core $C$: CODEC audio ||||||
| $G_{pb}$ | 20kHz | 20kHz | 640kHz | 80,000 | 1 |
| $f_c$ | 45kHz | 55kHz | 1.5MHz | 136,533 | 1 |
| $THD$ | 2kHz | 31kHz | 2.46MHz | 83,252 | 1 |
| Core $D$: Baseband down converter ||||||
| $IIP_3$ | 3.25MHz | 9.75MHz | 78MHz | 15,754 | 10 |
| $A_v$ | 26MHz | 26MHz | 26MHz | 9,228 | 4 |
| $DR$ | 26MHz | 26MHz | 26MHz | 31,508 | 4 |
| Core $E$: General purpose amplifier ||||||
| $SR$ | 69MHz | 69MHz | 69MHz | 5,400 | 5 |
| $A_v$ | 8MHz | 8MHz | 8MHz | 2,500 | 1 |

**Table 2.** Test requirements for the analog cores.

| # of wrappers | Wrapper sharing | $T(W)$ $W=32$ | $W=48$ | $W=64$ |
|---|---|---|---|---|
| 4 | {A,C} | 98.3 | **92.6** | 86.3 |
|   | {C,D} | 99.1 | **92.6** | 85.0 |
|   | {C,E} | 99.1 | **92.6** | 87.6 |
|   | {A,B} | **97.5** | **92.6** | **82.8** |
|   | {A,D} | 99.1 | 92.8 | 85.58 |
|   | {A,E} | 99.1 | **92.6** | 86.1 |
|   | {D,E} | 99.1 | 92.8 | 85.4 |
| 3 | {A,B,C} | 99.8 | 92.9 | 90.1 |
|   | {A,C,D} | 98.3 | 92.8 | 87.2 |
|   | {A,C,E} | 98.3 | **92.6** | 86.3 |
|   | {A,B,D} | 99.1 | 97.2 | 85.4 |
|   | {C,D,E} | 99.1 | 92.8 | 85.4 |
|   | {A,B,E} | 97.9 | 92.8 | **82.8** |
|   | {A,D,E} | 99.1 | 92.9 | 85.5 |
| 2 | {A,B,C,D} | 99.4 | 96.4 | 98.7 |
|   | {A,B,C,E} | 99.8 | 98.5 | 91.1 |
|   | {A,C,D,E} | 98.3 | 92.9 | 87.2 |
|   | {A,B,D,E} | 98.3 | 97.2 | 85.4 |
| 2 | {A,B,C}{D,E} | 99.8 | 94.9 | 90.1 |
|   | {A,C,D}{B,E} | 98.3 | 92.8 | 87.2 |
|   | {A,D,E}{B,C} | 98.3 | 92.9 | 87.8 |
|   | {C,D,E}{A,B} | 98.3 | **92.6** | 86.8 |
|   | {A,B,E}{C,D} | 97.9 | 92.8 | 86.8 |
|   | {A,C,E}{B,D} | 98.3 | 92.8 | 87.8 |
|   | {A,B,D}{C,E} | 99.1 | 97.2 | 85.4 |
| 1 | {A,B,C,D,E} | 100 | 100 | 100 |

**Table 3.** Test time results for SOC p93791m for different combinations of analog wrapper sharing.

For our experimental set-up, we have used a digital SOC from the ITC'02 SOC test benchmarks, namely p93791. We consider only p93791 because the test time for the other SOCs reaches a lower bound for relatively small values of TAM width. We have added five analog cores to the SOC. We refer to the mixed signal SOC as p93791m. The analog cores consist of a pair of baseband I-Q transmit path with a bandwidth of 500kHz, a CODEC audio path with a bandwidth of 50kHz, a baseband down conversion path, and a general purpose amplifier. These analog cores are taken from a commercial baseband cellular phone chip. The test set specifications for each of these analog cores are given in Table 2. Due to the lack of a standardized analog test generation tool, analog tests are defined manually based on the core specifications.

For the I-Q transmit path pair, six distinct specification-based tests are defined. These include the pass-band gain ($G_{pb}$), the cut-off frequency ($f_c$), the attenuation levels at 1MHz and 2MHz ($G_{1MHz}$ and $G_{2MHz}$), the third order input intercept ($IIP_3$), and the DC offset ($V_{off_{DC}}$), and the phase mismatch ($\Phi_{off}$). For the audio CODEC path, the specifications include $G_{pb}$, $f_c$, and the total harmonic distortion ($THD$). The Baseband down conversion path has three specified tests, namely a test for the $IIP_3$, a test for the gain ($A_v$), and a test for the dynamic range ($DR$). Lastly, the tests for the general purpose amplifier include a test for the slew rate ($SR$) and a test for the $A_v$. The TAM width requirements $w$ for each of the analog cores are also presented in Table 2. The self-test mode test time has not been considered for both analog and digital cores, thus, Table 2 presents the test time on the core-test mode only. It should be noted that the analog test wrappers are not limited to the tests listed in Table 2. The proposed test wrappers can be used for analog tests that are within the operating frequency and resolution of the data converters in the analog test wrappers.

Next, we study the impact of wrapper sharing among the analog cores on the overall test time of an SOC. Table 3 presents results for SOC p93791m. The test time is presented for all the combinations of analog wrapper sharing. The test times are normalized to the case of maximum test time, thus they are essentially the $C_T$ values for the combinations. As expected, the test time for the case when all the analog cores share the same wrapper results in the maximum test time. The combinations that result in the lowest test time are highlighted in Table 3. We conclude from the results that as the TAM width increases, the analog core combinations have a greater affect on the overall SOC test time. This is because with an increase in TAM width, the test time of the digital cores decreases and the test time of the analog cores becomes more prominent. Thus, the difference between the lowest and the highest test times of the various combinations for $W = 32, 48$, and 64 are 2.45, 7.36, and 17.18, respectively. It is also seen that the lowest test times are obtained for combinations with a lower degree of sharing. However, for $W = 48$ and $W = 64$, the lowest test times can also be obtained with combinations that have a high degree of sharing. These cases show that some test schedules can result in a low test test time, even with a high degree of sharing.

| $W$ | $C_{exh}$ | $\eta_{exh}$ | $S$ | $C$ | $\eta$ | $S$ | $\Delta\eta$ |
|---|---|---|---|---|---|---|---|
| \multicolumn{8}{c}{$w_T = 0.5, w_A = 0.5$} | | | | | | | |
| 32 | 71.9 | 26 | {A,B,E}{C,D} | 71.9 | 10 | {A,B,E}{C,D} | −61.5 |
| 40 | 69.7 | 26 | {A,B,D}{C,E} | 69.7 | 10 | {A,B,D}{C,E} | −61.5 |
| 48 | 69.3 | 26 | {C,D,E}{A,B} | 69.3 | 10 | {C,D,E}{A,B} | −61.5 |
| 56 | 68.9 | 26 | {A,B,E}{C,D} | 68.9 | 10 | {A,B,E}{C,D} | −61.5 |
| 64 | 65.7 | 26 | {A,B,D}{C,E} | 65.7 | 10 | {A,B,D}{C,E} | −61.5 |
| \multicolumn{8}{c}{$w_T = 0.8, w_A = 0.2$} | | | | | | | |
| 32 | 87.5 | 26 | {A,B,E}{C,D} | 88.4 | 7 | {A,C,D,E} | −73.0 |
| 40 | 84.0 | 26 | {A,B,D}{C,E} | 84.0 | 10 | {A,B,D}{C,E} | −61.5 |
| 48 | 83.3 | 26 | {C,D,E}{C,E} | 83.3 | 10 | {C,D,E}{C,E} | −61.5 |
| 56 | 82.7 | 26 | {A,B,E}{C,D} | 82.7 | 10 | {A,B,E}{C,D} | −61.5 |
| 64 | 77.5 | 26 | {A,B,D}{C,E} | 77.5 | 10 | {A,B,D}{C,E} | −61.5 |
| \multicolumn{8}{c}{$w_T = 0.2, w_A = 0.8$} | | | | | | | |
| 32 | 56.4 | 26 | {A,C,D}{B,E} | 56.4 | 10 | {A,C,D}{B,E} | −61.5 |
| 40 | 55.5 | 26 | {A,B,D}{C,E} | 55.5 | 10 | {A,B,D}{C,E} | −61.5 |
| 48 | 55.3 | 26 | {A,B,E}{C,D} | 55.3 | 10 | {A,B,E}{C,D} | −61.5 |
| 56 | 55.1 | 26 | {A,B,E}{C,D} | 55.1 | 10 | {A,B,E}{C,D} | −61.5 |
| 64 | 53.8 | 26 | {A,B,D}{C,E} | 53.8 | 10 | {A,B,D}{C,E} | −61.5 |

**Table 4.** Comparison of $Cost\_Optimizer$ with the exhaustive evaluation approach.

Table 4 presents the cost of sharing for a set of $\omega_T$ and $\omega_A$ values. The proposed $Cost\_Optimizer$ procedure is compared with the exhaustive evaluation approach described in Section 4. (Note that while exhaustive enumeration is possible for these test cases, the high CPU time notwithstanding, it is unlikely to be feasible for larger SOCs.) The $C_A$ values used are the same as those presented in Table 1. And the elimination criteria $\delta$ for the $Cost\_Optimizer$ approach is chosen to be zero. Recall that the exhaustive evaluation approach always results in optimal results, although at the expensive of greater computation time. It is seen that the $Cost\_Optimizer$ procedure also gives optimal results for all but one case with a much lower computation time. In Table 4, $\eta_{exh}$ and $\eta$ represent the number of combinations evaluated to arrive at the results, and $S$ represents the combination of core sharing selected. $\eta_{exh}$ is always 26, since there are a total of 26 combinations. The lower bound on $\eta$ is 4, since the best combinations of four groups have to be evaluated. It is seen that the reduction in the number of combinations evaluated is significant even for the cases where $Cost\_Optimizer$ yields optimal results. The percentage reduction in the number of evaluations $\Delta\eta$ is also reported ($\Delta\eta = \frac{\eta - \eta_{exh}}{\eta_{exh}} \times 100$). On an average, the $Cost\_Optimizer$ procedure takes 6 minutes to complete on a SunW Ultra 5_10, whereas the exhaustive approach requires approximately 20 minutes to complete.

## 7 Conclusion

We have presented a resource optimization technique and a cost-oriented optimization heuristic to reduce the overall test cost of mixed-signal SOCs. In the resource optimization approach, we show that multiple analog cores can share analog test wrappers to reduce the area overhead. The cost-oriented optimization approach uses a well known TAM optimization approach together with the analog wrapper optimization technique to give a cost efficient TAM architecture and test schedule for a mixed-signal SOC. We have also presented transistor-level simulation results to demonstrate the feasibility of the analog test wrappers. We have presented experimental results demonstrating that the test cost can be reduced significantly, using the proposed optimization techniques. As part of future work, we are studying ways of refining the cost measure based on the knowledge of core placement. We are investigating the cost of testing the data converters in the analog test wrappers. We are working with industrial partners to apply this method to real-life mixed-signal SOCs.